\documentclass[conference]{IEEEtran}
\IEEEoverridecommandlockouts
\usepackage{cite}
\usepackage[numbers,sort&compress]{natbib}
\usepackage[utf8]{inputenc} 
\usepackage[T1]{fontenc}
\usepackage{url}
\usepackage{ifthen}
\usepackage[cmex10]{amsmath}
\usepackage{array} % Use the [cmex10] option to ensure complicance
\usepackage{enumerate}
\usepackage{amssymb}
\usepackage{amsmath} 
\usepackage[lined,boxed,commentsnumbered, ruled]{algorithm2e}
\usepackage{mathrsfs}
\usepackage{bm}
\usepackage{float}
\usepackage{tikz}
\usepackage{algorithmic}
\usetikzlibrary{arrows}
\usepackage{subfigure}
\usepackage{graphicx,booktabs,multirow}
\usepackage{epstopdf}
\usepackage{subfigure}
\usepackage{multirow}
\usepackage{tabularx} 
\usepackage{booktabs}
\usepackage{stmaryrd}
\usepackage{tikz}
\usepackage{pgfplots}
\usepackage{textcomp}
\usepackage{siunitx}
\pgfplotsset{compat=1.12}

\definecolor{colorhkust}{RGB}{20,43,140}
\definecolor{colortsinghua}{RGB}{116,52,129}
\definecolor{color1}{RGB}{128,0,0}

\usepackage{amsthm}

\newtheorem{proposition}{Proposition}
\theoremstyle{definition}

\renewcommand{\baselinestretch}{0.945}

\newcommand{\diagg}{\mathrm{diag}}

\def\BibTeX{{\rm B\kern-.05em{\sc i\kern-.025em b}\kern-.08em
		T\kern-.1667em\lower.7ex\hbox{E}\kern-.125emX}}
\begin{document}

\title{Intelligent Reflecting Surface for Downlink Non-Orthogonal Multiple Access Networks}

\author{\IEEEauthorblockN{Min Fu, Yong Zhou, and Yuanming Shi}\\
        
        \IEEEauthorblockA{
                School of Information Science and Technology,
                ShanghaiTech University, Shanghai 201210, China\\
                E-mail: \{fumin, zhouyong, shiym\}@shanghaitech.edu.cn
        }
        
}

\maketitle
\begin{abstract}
%Non-orthogonal multiple access (NOMA) has been envisioned as one of the key enabling techniques to satisfy the requirements for 5G wireless communications, such as high spectral efficiency and massive connectivity.
Intelligent reflecting surface (IRS) has recently been recognized as a promising technology to enhance the energy and spectrum efficiency of wireless networks by controlling the wireless medium with the configurable electromagnetic materials. 
In this paper, we consider the downlink transmit power minimization problem for a  IRS-empowered non-orthogonal multiple access (NOMA) network by jointly optimizing the transmit beamformers at the BS and the phase shift matrix at the IRS. 
However, this problem turns out to be a highly intractable non-convex bi-quadratic programming problem, for which an alternative minimization framework is proposed via solving the non-convex quadratic programs alternatively. We further develop a novel difference-of-convex (DC) programming algorithm to solve the resulting non-convex quadratic programs efficiently by lifting the quadratic programs into rank-one constrained matrix optimization problems, followed by representing the non-convex rank function as a DC function. Simulation results demonstrate  the performance gains of the proposed method. 

\end{abstract}

\section{Introduction}
Intelligent reflecting surface (IRS), as an emerging cost-effective technology, has the great potential to significantly enhance the spectrum and energy efficiency of wireless networks by reconfiguring the wireless propagation environment \cite{di2019smart,liang2019large}. 
Specifically, an IRS is a metasurface composed of a large number of passive reflecting elements, each of which is able to independently change the phase shift of the incident signal to be reflected \cite{Akyildiz_ComMag18metasurface, huang2018large}. 
By adaptively altering the reflected signal propagation, an IRS is able to achieve desired channel responses for constructive signal combination and interference cancellation at the receivers, thereby enhancing the wireless network performance \cite{Rui_arXiv19IRSmag}. 

The beamforming design for IRS-empowered wireless networks has recently attracted considerable attention \cite{wu2018intelligent, nadeem2019large, guo2019weighted, jiang2019over}. 
The base station (BS) transmit power minimization problem was considered in  \cite{wu2018intelligent} by jointly optimizing active beamforming at the BS and passive beamforming at the IRS. 
It was demonstrated  that the IRS can significantly reduce the energy consumption in wireless networks \cite{wu2018intelligent}. The achievable maximin data rate optimization problem was considered in \cite{nadeem2019large} via random matrix theory, while the fractional programming algorithm was developed in \cite{guo2019weighted} to solve the weighted sum-rate maximization problem. In addition, the IRS was leveraged to boost the received signal power for over-the-air computation in multiple access networks \cite{jiang2019over}. 

Non-orthogonal multiple access (NOMA) becomes one of the key enabling techniques in wireless networks to support massive connected devices and enhance the spectral efficiency \cite{ding2015application, L2015NOMA}. 
With power-domain NOMA, a BS can concurrently serve multiple users in the same resource block by using superposition coding and performing successive interference cancellation at the BS and users, respectively \cite{Islam_ST17NOMA}. The performance gains of NOMA over orthogonal multiple access  has been demonstrated in massive multiple input single output networks \cite{alavi2018beamforming}, millimeter wave networks \cite{zhou2018coverage}, and etc. In the emerging 6G networks, it becomes critical to support new intelligent services with stringent requirements on data rates, latency, and connectivity \cite{Letaeif_6GMag19,Ertugrul2019wireless}, for which we shall propose an IRS-empowered NOMA technique to provide a potential multiple access solution for the future 6G networks.

 In this paper, we consider a downlink IRS-empowered NOMA network, where a single BS serves multiple users with the help of an IRS. Specifically, we propose to jointly optimize
the beamforming vectors at the BS and the phase shift matrix at the IRS to minimize the total transmit power consumption at the BS, while satisfying the quality-of-service (QoS) requirements of each user. 
The formulated problem turns out to be a highly intractable non-convex bi-quadratic programming problem, for which we present an alternative optimization method to update the beamforming vectors and
the phase shift matrix alternatively. We further propose to lift the non-convex quadratically constrained
quadratic programming (QCQP) problems in the alternatively updating procedure into rank-one constrained matrix optimization problems. Although dropping the fixed--rank constraint using the semidefinite relaxation (SDR) \cite{ma2010semidefinite} approach could yield a convex problem \cite{alavi2018beamforming}, the obtained solutions normally yield poor performance in the high-dimensional settings \cite{yang2018federated, jiang2019over}.

%To overcome the limitation of the SDR method, we in this paper propose an alternating difference-of-convex (DC) method
%by recasting the non-convex QCQP problem as a difference-of-convex (DC) programming problem. 

To overcome the limitations of the SDR method, we instead propose a difference-of-convex (DC) method by recasting the non-convex QCQP problem as a DC programming problem. 
Specifically, we present an  exact DC representation for the non-convex rank-one constrained positive semidefinite (PSD) matrix by exploiting the difference between the trace norm and the spectral norm \cite{Yuanming-TSP19WDC}.  
We then develop an efficient DC algorithm  to solve the resulting non-convex DC programming problem. 
Due to the superiority of the proposed DC representation, the numerical results show that the proposed  DC method considerably outperforms the existing methods in terms of minimizing the downlink transmit power in the IRS-empowered NOMA network.

\section{System Model and Problem Formulation}

\subsection{System Model}
\begin{figure}
        \centering
        \includegraphics[width=0.9\columnwidth,height=3.5cm]{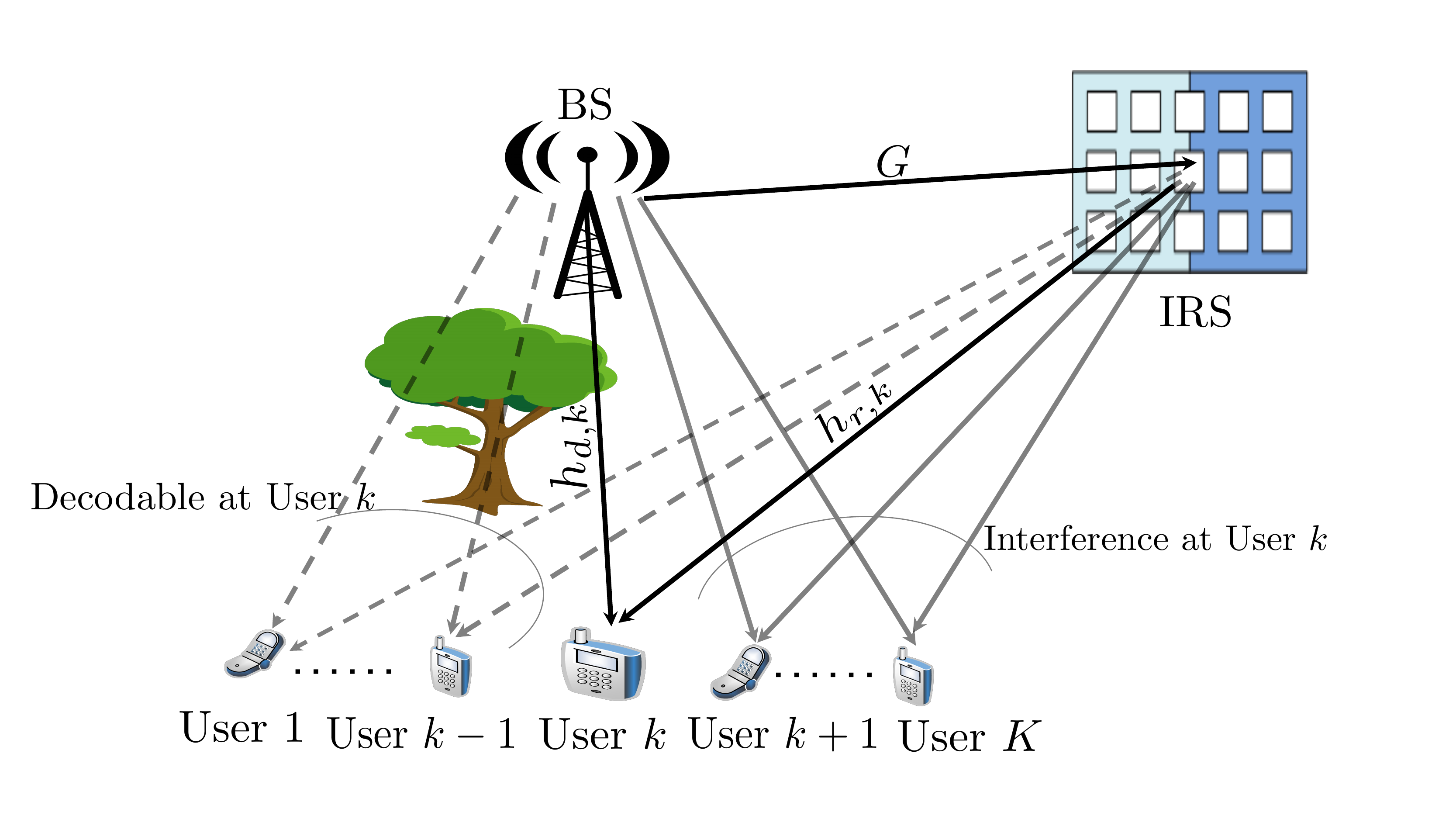} 
        \vspace{-3mm}
        \caption{An IRS-empowered single-cell NOMA network.}\label{systemodel}
        \vspace{-6mm}
\end{figure}

Consider the  downlink NOMA transmission of a single-cell network, where an IRS with $N$ passive reflecting elements is deployed to assist the transmission from a $M$-antenna BS to $K$ single-antenna users, as shown in Fig.\ref{systemodel}. 
We consider a dense scenario, where the number of users is not less than the number of antennas at the BS, i.e., $K \ge M$. 
We denote $s_k$ and $ \bm w_k\in\mathbb{C}^{M} $ as the signal and beamforming vector for user $U_k$, respectively, where $k \in \mathcal{K} = \{1, 2, \ldots, K \}$. 
Without loss of generality, signal $s_k$ is assumed to have zero mean and unit variance,  i.e., $\mathbb{E}[s_ks_k^{\sf H}] = 1, \forall \, k \in \mathcal{K}$, where $(\cdot)^{\sf H}$  denotes the conjugate transpose. 
After transmitted by the BS and reflected by the IRS, the signal received at user $U_k$ is given by 
\setlength\arraycolsep{2pt}
\begin{eqnarray}\label{receive1}
y_k = (\bm h_{r,k}^{\sf{H}}\bm \Theta\bm G+\bm h_{d,k}^{\sf H})\sum_{j=1}^{K}\bm w_js_j+e_k, \forall \, k \in \mathcal{K},
\end{eqnarray}
where $ \bm h_{d,k}\in\mathbb{C}^M, \bm G\in\mathbb{C}^{N\times M}$, and $\bm h_{r,k}\in\mathbb{C}^{N} $ denote the channel responses from the BS to  $U_k$, from the BS to the IRS, and from the IRS to  $U_k$, respectively, $ \bm\Theta={\diagg}(e^{j\theta_1},\ldots,e^{j\theta_N}) \in \mathbb{ C}^{N \times N}$ represents the diagonal phase shift matrix of the IRS with $  \theta_n\in[0,2\pi], n \in \{1, \ldots, N\} $,  and $ e_k\sim\mathcal{CN}(0,\sigma^2) $ is the additive white Gaussian noise (AWGN).

To facilitate NOMA transmission, the users are ordered based on their channel quality with respect to the BS, which has been widely adopted in the literature. 
Specifically, $K$ users are ordered as $\|\bm h_{d,1}\|_2 \leq \|\bm h_{d,2}\|_2 \leq \ldots \leq\|\bm h_{d,K}\|_2$, where $\|\bm h_{d,k}\|_2$ denotes the channel quality between user $U_k$ and the BS.  
Based on the decoding principle of NOMA, user $U_1$ directly decodes its own signal by treating the signals intended for other users as noise. 
On the other hand, user $U_k, k \ge 2$, sequentially decodes and removes other users' signals, until its own signal $s_k$ is decoded. 
%Note that this ordering may not be optimal. However, our work in this paper does not focus on the optimal decoding ordering problem.
%Under the ordering based on distance, the $U_1$ is the weakest (and hence cannot decode any interfering signals), while $U_K$ is the strongest user,
%and is able to nullify all other users interference by performing SIC. The other users are placed in an increasing order with respect to their index numbers. Base on this ordering,  user $U_k$ can detect and remove the first $k-1$ users' signals in a successive manner whereas the message of the other users, i.e., from $U_{k+1}$ to $U_K$, is treated as noise. 
Thus, the signal at user $U_l$ before decoding signal $s_k$ can be expressed as 
\begin{eqnarray}\label{receive2}
y_l^k = (\bm h_{r,l}^{\sf{H}}\bm \Theta\bm G+\bm h_{d,l}^{\sf H})\sum_{j=k}^{K}\bm w_js_j+e_l,  \ \forall \, l=k,\ldots, K.
\end{eqnarray}

After successfully canceling the signals intended for users $\{U_1, \ldots, U_{k-1}\}$, the achievable data rate of  $U_k$ is given by
\begin{eqnarray}\label{achievable rate}
R_k = \text{log}_2 \left(1+ \mathop{\text{min}}_{l\in[k,K]} \text{SINR}_l^k \right),
\end{eqnarray}
where  the signal-to-interference-plus-noise ratio (SINR) of signal $s_k$ at user $U_l$ can be expressed as 
\begin{eqnarray}\label{SINR}
\text{SINR}_l^k = \frac{|(\bm h_{r,l}^{\sf{H}}\bm \Theta\bm G+\bm h_{d,l}^{\sf H})\bm w_k|^2}{ |(\bm h_{r,l}^{\sf{H}}\bm \Theta\bm G+\bm h_{d,l}^{\sf H})\sum_{j=k+1}^{K}\bm w_j |^2+ \sigma^2}.
\end{eqnarray}

\subsection{Problem Formulation}
In this subsection, we formulate an optimization problem to jointly optimize the beamforming vectors (i.e., $\{\bm w_k, \forall \, k \in \mathcal{K}\}$) at the BS and the phase shift matrix (i.e., $\bm \Theta$) at the IRS, aiming to minimize the total transmit power while satisfying the data rate requirements of each user. 
The transmit power minimization problem is thus formulated as  
\begin{eqnarray}\label{mixed}
\mathscr{P}_1:\mathop{\text{minimize}}_{ \{\bm w_k\},\bm \Theta}
&&\sum_{k=1}^{K}\|\bm w_k\|^2 \nonumber\\
\text{subject to}&& \text{log}_2\left(1+ \mathop{\text{min}}_{l\in[k,K]} \text{SINR}_l^k \right) \geq R_k^{\text{min}},\forall \,k , \\
\label{eqn_theta}&&0\le \theta_n\le 2\pi, \forall \, n,
\end{eqnarray}
where %$\{\bm w_k\}_{1\leq k\leq K}\in \mathbb{ C}^{M}$ and $\bm \Theta \in \mathbb{ C}^{N \times N}$  are optimization parameters, and 
$\|\bm w_k\|^2$ is the power assigned to transmit signal $s_k$ and 
$R_k^{\text{min}}$ denotes the minimum data rate requirement of user $U_k$. 
To assist the algorithm design, we rewrite 
%we make some equivalent transformations to simplify the constraints in problem $\mathscr{P}_1$. 
constraints  $\eqref{mixed}$ of problem $\mathscr{P}_1$ as
\begin{eqnarray}\label{achievable rate constraint}
\mathop{\text{min}}_{l\in[k,K]} \text{SINR}_l^k \geq \gamma_k^{\text{min}}, \forall \, k,
\end{eqnarray}
where $\gamma_k^{\text{min}} = 2^{R_k^{\text{min}}} -1$ is the minimum SINR required to successfully decode signal $s_k$. Constraints $\eqref{achievable rate constraint}$ can be further rewritten as
\begin{eqnarray}\label{simplyrate}
 &&\gamma_k^{\text{min}} \Big(\big|(\bm h_{r,l}^{\sf{H}}\bm \Theta\bm G+\bm h_{d,l}^{\sf H})\sum_{j=k+1}^{K}\bm w_j\big|^2 + \sigma^2\Big) \nonumber \\
&& \ \ \ \ \  \leq |(\bm h_{r,l}^{\sf{H}}\bm \Theta\bm G+\bm h_{d,l}^{\sf H})\bm w_k|^2,
  \forall \, k, l = k,\cdots, K.
\end{eqnarray}
Therefore, problem $\mathscr{P}_1$ can be equivalently reformulated as
\begin{eqnarray}\label{mixedSINR}
\mathscr{P}_2:\mathop{\text{minimize}}_{ \{ \bm w_k\},\bm \Theta}
&&\sum_{k=1}^{K}\|\bm w_k\|^2\nonumber \\
\text{subject to}&&   \mathrm{constraints}\; \eqref{eqn_theta}, \eqref{simplyrate}. 
\end{eqnarray}

However, problem $\mathscr{P}_2$ is still highly intractable due to the non-convex bi-quadratic constraints (i.e., \eqref{simplyrate}), in which the beamforming vectors and the phase shift matrix are coupled. To address this challenge,  we present an alternating optimization approach to solve problem $\mathscr{P}_2$ in the next section.

\section{Alternating Optimization Framework}
 In this section, we present an alternating minimization approach  to solve problem $\mathscr{P}_2$ with good performance. 
In particular, the beamforming vectors $\{\bm w_k, k \in \mathcal{K}\}$ and phase shift matrix $\bm \Theta$ are optimized alternatively until convergence. 
%, i.e., optimizing one parameter while keeping the other one fixed in each iteration. 
Moreover, we transform the resulting non-convex QCQP problem in the procedure of alternating minimization into a rank-one constrained matrix optimization problem via matrix lifting. 

% \subsection{Alternating Optimization}
For a given phase shift matrix $ \bm\Theta $, channel response $\bm h_l^{\sf H}= \bm h_{r,l}^{\sf{H}}\bm \Theta\bm G+\bm h_{d,l}^{\sf H} \in \mathbb{ C}^{1\times M}$ is fixed, and hence problem $ \mathscr{P}_2 $ can be simplified as the following non-convex QCQP problem
\begin{eqnarray}\label{fixtheta}
\mathop{\text{minimize}}_{\{ \bm w_k\}}
&&\sum_{k=1}^{K}\|\bm w_k\|^2\nonumber \\
\text{subject to}&&  \gamma_k^{\text{min}} \left(\sum_{j=k+1}^{K}|\bm h_{l}^{\sf{H}}\bm w_j|^2 + \sigma^2\right) \leq |\bm h_{l}^{\sf{H}}\bm w_k|^2 ,\nonumber \\
&&\ \ \ \  \ \ \ \ \ \ \ \ \ \forall k, l = k,\ldots, K.
\end{eqnarray}

To address the non-convex constraints in  problem $\eqref{fixtheta}$, a natural way is to reformulate problem $\eqref{fixtheta}$ as a semidefinite programming (SDP) problem by using the matrix lifting technique \cite{alavi2018beamforming}.
By lifting vector $\bm w_k$ into a PSD matrix $\bm W_k = \bm w_k \bm w_k^\mathsf{H} \in \mathbb{ C}^{M\times M}$ with $\mathrm{rank}(\bm W_k)=1$, $\forall \, k \in \mathcal{K}$, problem $\eqref{fixtheta}$ can be equivalently rewritten as 
\begin{eqnarray}\label{liftomega}
\mathop{\text{minimize}}_{ \{ \bm W_k\}}
&&\sum_{k=1}^{K}\text{Tr}(\bm W_k)\nonumber \\
\text{subject to}&&  \gamma_k^{\text{min}} \left(\sum_{j=k+1}^{K}\text{Tr}(\bm H_l\bm W_j) + \sigma^2\right) \nonumber\leq \text{Tr}(\bm H_l\bm W_k),\nonumber \\
&&\ \ \ \ \ \ \ \ \ \ \ \ \ \ \ \forall \, k, l = k,\ldots, K,\nonumber\\
&&\bm W_k \succcurlyeq 0, \mathrm{rank}(\bm W_k) = 1, \forall \, k,
\end{eqnarray}
where $\bm H_l = \bm h_l \bm h_l^\mathsf{H}\in \mathbb{ C}^{M\times M}$. 
%Note that beamforming vector $\bm w_k$ can be extracted from the optimal solution $\bm W_k^*$, $\forall \, k \in \mathcal{K}$.

On the other hand, for given beamforming vectors $ \{\bm w_k, \forall \, k \in \mathcal{K}\}$,  we denote $b_{l,k} = \bm h_{d,l}^{\sf H}\bm w_k$, $\forall \, k, l= k,\ldots, K$,  $ v_n=e^{-j\theta_n}, \forall \, n= 1,\ldots, N $, and $\bm v^{\sf H}\bm a_{l,k} = \bm h_{r,l}^{\sf H}\bm \Theta{\bm G}\bm w_k$, where $ \bm v=[e^{j\theta_1},\ldots,e^{j\theta_N}] ^{\sf H}$ and $ \bm a_{l,k} = {\diagg}(\bm h_{r,l}^{\sf H})\bm G\bm w_k $.  
As a result,  problem $ \mathscr{P}_2 $ can be simplified into the following non-convex feasibility detection problem
\begin{eqnarray}\label{fixomega}
\mathop{\text{find}}
&&{\bm v}\nonumber \\
\text{subject to}&& \gamma_k^{\text{min}} \left(\sum_{j=k+1}^{K}|\bm v^{\sf H} \bm a_{l,j}+ b_{l,j}|^2 + \sigma^2\right) \nonumber\leq \nonumber \\
&&\ \ \ \ \ \ \ \ \ \ |\bm v^{\sf H} \bm a_{l,k}+ b_{l,k}|^2,  \forall \, k, l = k,\ldots, K,\nonumber\\
&&|\bm v_n| =1 , \forall \, n = 1,\ldots, N.
\end{eqnarray}

Although problem $\eqref{fixomega}$ is non-convex and inhomogeneous, it can be reformulated as a homogenous non-convex QCQP problem by introducing an auxiliary variable $t$. 
Thus, problem $\eqref{fixomega}$ can be rewritten as 
\begin{eqnarray}\label{fixomega1}
\mathop{\text{find}}
&&{\tilde{\bm v}}\nonumber \\
\text{subject to}&&  \gamma_k^{\text{min}} \left(\sum_{j=k+1}^{K}\tilde{\bm v}^{\sf H} \bm R_{l,j}\tilde{\bm v} +  b_{l,j}^2 +\sigma^2\right) \nonumber\leq  \nonumber \\
&&\ \ \ \ \ \ \ \ \ \ \tilde{\bm v}^{\sf H} \bm R_{l,k}\tilde{\bm v}+ b_{l,k}^2, \forall \, k, l = k,\ldots, K,\nonumber\\
&&|\tilde{\bm v}_n| =1 , \forall \, n = 1,\ldots,N+1,
\end{eqnarray}
where
\begin{eqnarray}
 \bm R_{l,k} = 
\begin{bmatrix}
        \bm a_{l,k}\bm a_{l,k}^{\sf H}     &    \bm a_{l,k}     b_{l,k}      \\
        b_{l,k}^{\sf H}\bm a_{l,k}^{\sf H}       & 0 
\end{bmatrix}, \tilde{\bm v}=
\begin{bmatrix}
\bm v    \\
t
\end{bmatrix}.
\end{eqnarray}

If $ \tilde{\bm v}^*$ is a feasible solution to problem $\eqref{fixomega1}$, then we obtain a feasible solution to problem $\eqref{fixomega}$ by setting $\bm{v} =   [\tilde{\bm v}^*/\tilde{\bm v}^*_{N+1}]_{(1:N)} $, where $[\bm x]_{(1:N)}$ denotes the first $N$ elements of vector $\bm x$.

Similarly, we adopt the matrix lifting technique to reformulate the non-convex quadratic constraints in problem $\eqref{fixomega1}$.  By denoting $\bm V = \tilde{\bm v}\tilde{\bm v}^{\sf H}$ and $\text{Tr}(\bm R_{l,k}\bm V) = \tilde{\bm v}^{\sf H} \bm R_{l,k}\tilde{\bm v}$, problem $\eqref{fixomega1}$ can be equivalently rewritten as the following rank-one constrained matrix optimization problem: 
\begin{eqnarray}\label{liftv}
\mathop{\text{find}}
&&{\bm V}\nonumber \\
\text{subject to}&&  \gamma_k^{\text{min}} \left(\sum_{j=k+1}^{K}\text{Tr}(\bm R_{l,j}\bm V) + b_{l,j}^2+ \sigma^2\right) \nonumber\leq \nonumber \\
&&\ \ \ \ \ \ \ \ \ \ \text{Tr}(\bm R_{l,k}\bm V) + b_{l,k}^2, \forall \, k, l = k,\ldots, K,\nonumber\\
&&\bm V_{n,n} =1 , \forall \, n= 1,\ldots,N+1,\nonumber\\
&&\bm V \succcurlyeq 0, \mathrm{rank}(\bm V) = 1.
\end{eqnarray}

Problems $\eqref{liftomega}$ and $\eqref{liftv}$ are still non-convex due to the low-rank constraints.
%Although problems , their algorithmic advantages can be exploited to develop efficient algorithms. 
%\subsection{Problem Analysis}
%Both  problems $\eqref{liftomega}$ and  $\eqref{liftv}$ are still non-convex optimization problems due to rank-one  constraints. 
%Convex relaxation is a well-known solution to deal with the non-convex fixed-rank constraint.
The SDR technique \cite{ma2010semidefinite} can be used to  deal with the non-convex rank constrains, as the fixed-rank constrained SDP problem after dropping the rank-one constraint can be  solved by the existing solvers. 
If the returned solution of the relaxed SDP problem fails to be rank-one, then Gaussian randomization \cite{ma2010semidefinite} is adopted to obtain a suboptimal solution. 
Although the SDR technique can solve problems $\eqref{liftomega}$ and $\eqref{liftv}$, the probability of the obtained solution being rank-one is small, especially when the dimension of the optimization parameters is high \cite{yang2018federated, jiang2019over}. 

To address the limitations of the SDR technique, we  shall propose an exact DC representation for the rank constraint of the PSD matrix by exploiting the difference between the trace norm and the spectral norm in the following section.

%This is achieved by exploiting the difference between the trace norm and the spectral norm for the PSD matrix rank function. 
%Subsequently, we develop an efficient DC algorithm  to solve the resulting non-convex DC programming problem by solving successively solving the convex relaxation of primal problem and  dual problem of DC programming. 

%{\color{red}Due to the superiority of the proposed DC representation, the numerical results will show that the proposed alternating DC approach considerably outperforms the existing convex methods in terms of the total transmit power.}

\section{Proposed Alternating DC Method}
%we propose an alternating DC method. 
%In this section, we propose an alternating DC method to solve problem $\mathscr{P}_2$.  Instead of solving problems $\eqref{liftomega}$ and $\eqref{liftv}$, we optimize two DC programming problems alternatively.
In this section, we present an exact DC representation for  the rank function, followed by proposing an  alternating DC method to solve the
original fixed-rank constrained matrix optimization problem.

 %resulted non-convex DC programming problem.
% by successively solving the convex relaxation of primal and dual problems of DC programming. 

\subsection{Proposed Alternating DC Programming}
Firstly, we introduce an exact DC representation for the fixed-rank constraint in the following proposition.
\begin{proposition} 
For PSD matrix $\bm X\in \mathbb{C}^{N\times N}$ and $\mathrm{Tr}(\bm X) >0$, we have \cite{yang2018federated}
        \begin{equation*}
        \mathrm{rank}(\bm X)=1\Leftrightarrow \mathrm{Tr}(\bm X)-\|\bm X\|_2=0,
        \end{equation*}
        where  trace norm $\mathrm{Tr}(\bm X)=\sum_{i=1}^{N} \sigma_i(\bm X)$ and spectral norm $\|\bm X\|_2=\sigma_1(\bm X)$ with $ \sigma_i(\bm X) $ denoting the $ i $-th largest singular value of matrix $ \bm X $.
\end{proposition}

We then apply the DC framework to problems $\eqref{liftomega}$ and $\eqref{liftv}$. 
Given the phase shift matrix $\bm \Theta$, we solve the following DC programming to find $K$ rank-one matrices to problem $\eqref{liftomega}$: 
\begin{eqnarray}\label{liftomegaDC}
\mathop{\text{minimize}}_{ \{ \bm W_k\}}
&&\sum_{k=1}^{K}\text{Tr}({ \bm W}_k) + \rho\sum_{k=1}^{K}\Big(\text{Tr}(\bm W_k)-\|\bm W_k\|_2\Big)\nonumber \\
\text{subject to}&&  \gamma_k^{\text{min}} \left(\sum_{j=k+1}^{K}\text{Tr}({\bm H}_l^\mathsf{H}\bm W_j) + \sigma^2\right) \nonumber\leq \text{Tr}(\bm H_l^\mathsf{H}\bm W_k),\nonumber \\
&&\ \ \ \ \ \ \ \ \ \ \ \ \forall \, k, l = k,\ldots, K,\nonumber\\
&&\bm W_k \succcurlyeq 0, \forall \, k,
\end{eqnarray}
where $\rho>0$ is a penalty parameter. By enforcing the penalty term to be zero, problem $\eqref{liftomegaDC}$ induces $K$ rank-one matrices. 
After solving problem \eqref{liftomegaDC}, we can recover the beamforming vectors $\bm w_k, k \in \mathcal{K}$, through Cholesky decomposition $\bm W_k^* = \bm w_k \bm w_k^{\sf H}$.

Similarly, given beamforming vectors $\{\bm w_k, k \in \mathcal{K} \}$, we  minimize the following difference between the trace norm and the spectral norm to detect the feasibility of problem $\eqref{liftv}$:  
\begin{eqnarray}\label{liftvDC}
\mathop{\text{minimize}}_{\bm V}
&&\text{Tr}(\bm V)-\|\bm V\|_2\nonumber \\
\text{subject to}&&  \gamma_k^{\text{min}}\left (\sum_{j=k+1}^{K}\text{Tr}(\bm R_{l,j}\bm V) + b_{l,j}^2+ \sigma^2\right) \nonumber\leq  \nonumber \\&& \ \ \ \ \ \ \ \ \ \text{Tr}(\bm R_{l,k}\bm V) +b_{l,k}^2, \forall \, k, l = k,\ldots, K,\nonumber\\
&&\bm V_{n,n} =1 , \forall \, n = 1,\ldots,N+1,\nonumber\\
&&\bm V \succcurlyeq 0.
\end{eqnarray}
Specifically, when the objective value of problem $\eqref{liftvDC}$ becomes zero, we  obtain an exact rank-one optimal solution, denoted as $\bm V^*$. 
Using Cholesky decomposition $\bm V^* = \tilde{\bm v}  \tilde{\bm v}^{\sf H}$, we  obtain a feasible solution $\tilde{\bm v} $ to problem $\eqref{fixomega1}$. If the objective value fails to be zero, we claim that the original problem $\eqref{fixomega}$ is infeasible. 

\subsection{DC Algorithm for Problems $\eqref{liftomegaDC}$ and $\eqref{liftvDC}$ }
Although the DC programming problems $\eqref{liftomegaDC}$ and $\eqref{liftvDC}$ are still non-convex, they have a good structure, which can be exploited to develop efficient algorithms by successively solving the convex relaxation versions of the primal and dual problems of DC programming \cite{Dinh1997d.c}.  
Specifically, we can equivalently rewrite problem $\eqref{liftomegaDC}$ as
\begin{eqnarray}
\mathop{\text{minimize}}_{ \{ \bm W_k\}}
&&\sum_{k=1}^{K}\text{Tr}({ \bm W}_k) + \rho\sum_{k=1}^{K}\Big(\text{Tr}(\bm W_k)-\|\bm W_k\|_2\Big) +\nonumber\\ && \ \  \ \ 
\ \ \ \ \  \mathit{I}_{\mathcal{C}_1}(\{\bm W_k\}),
\end{eqnarray}
and problem $\eqref{liftvDC}$ as
\begin{eqnarray}
\mathop{\text{minimize}}_{\bm V}
&&\text{Tr}(\bm V)-\|\bm V\|_2 + \mathit{I}_{\mathcal{C}_2}(\bm{V}),
\end{eqnarray}
where $\mathcal{C}_1$, $\mathcal{C}_2$  are PSD cones that respectively satisfy the constraints in problems $\eqref{liftomegaDC}$ and $\eqref{liftvDC}$ and the indicator function is defined as
$$\mathit{I}_{\mathcal{C}}(\bm{Z})=
\begin{cases}
0,& \bm{Z}\in \mathcal{C}\\
+\infty,& \text{otherwise}
\end{cases},$$

It turns out that problems $\eqref{liftomegaDC}$ and $\eqref{liftvDC}$ have the structure of minimizing the difference of two convex functions, i.e., 
\begin{eqnarray}\label{DC structure}
\mathop{\textrm{minimize}}_{ \bm{Z}\in\mathbb{ C}^{m\times n} } &&f = g(\bm{Z}) - h(\bm{Z}), 
\end{eqnarray}  

 According to the Fenchel's duality \cite{rockafellar2015convex}, the dual problem of problem $\eqref{DC structure}$ is represented by
\begin{eqnarray}\label{dual DC structure}
\mathop{\textrm{minimize}}_{ \bm{Y}\in\mathbb{ C}^{m\times n}} &&h^*(\bm{Y}) - g^*(\bm{Y}),
\end{eqnarray}  
where   $g^*$ and $h^*$ are the conjugate functions of $g$ and $h$, respectively. The conjugate function is defined as
\begin{eqnarray}\label{conjugate func}
h^*(\bm{Y}) = \mathop{\textrm{sup}}_{\bm{Y}\in \mathbb{ C}^{m\times n}}  \{ \langle \bm{Z}, \bm{Y}\rangle - h(\bm{Z}) : \bm Z \in \mathcal{Z} \},
\end{eqnarray}  
where the inner product is defined as  $\langle \bm{X}, \bm{Y}\rangle = \mathfrak{R}(\text{Tr}(\bm{X}^\mathsf{H}\bm{Y}))$ according to Wirtinger's calculus \cite{Shi_TSP18demixing} in the complex domain. The DC algorithm iteratively updates both primal and dual variables via successive convex approximation. The  $t^{th}$ iteration is given by
\begin{align}\label{t-iteration}
\bm{Y}^t &= \mathop{\textrm{arg\,inf}}_{ \bm{Y}}\  h^*(\bm{Y})-\big[g^*(\bm{Y}^{t-1}) + \langle \bm{Y}-\bm{Y}^{t-1}, \bm{Z}^t\rangle\big],\\
\bm{Z}^{t+1} &= \mathop{\textrm{arg\,inf}}_{ \bm{Z}}\  g(\bm{Z})-\big[h(\bm{Z}^{t}) + \langle \bm{Z}-\bm{Z}^{t}, \bm{Y}^t\rangle\big]. 
\end{align}     
Based on the Fenchel biconjugation theorem \cite{rockafellar2015convex}, $\eqref{t-iteration}$ can be represented as
\begin{align}\label{subgradient}
&\bm{Y}^t  \in \partial_{\bm{Z}^t}h,
\end{align}     
where $\partial_{\bm{Z}^t}h$ is the sub-gradient of $h$ with respect to $\bm{Z}$ at $\bm{Z}^t$. 
Thus, $\{\bm W_k^{t}, \forall \, k \in \mathcal{K}\}$ at the $t^{th}$ iteration for problem $\eqref{liftomegaDC}$ are the solution to the following convex optimization problem 
\begin{eqnarray}\label{subliftomegaDC}
\mathop{\text{minimize}}_{ \{ \bm W_k\}}
&&\sum_{k=1}^{K}\text{Tr}({ \bm W}_k) + \rho\sum_{k=1}^{K}\Big(\langle \bm{W_k},\bm I -\partial_{\bm{W_k}^{t-1}}\|\bm{W}_k\|_2\rangle\Big)\nonumber \\
\text{subject to}&&  \gamma_k^{\text{min}} \left(\sum_{j=k+1}^{K}\text{Tr}({\bm H}_l^\mathsf{H}\bm W_j) + \sigma^2\right) \nonumber\leq \text{Tr}(\bm H_l^\mathsf{H}\bm W_k),\nonumber \\
&&\ \ \ \ \ \ \forall \, k, l = k,\ldots, K,\nonumber\\
&&\bm W_k \succcurlyeq 0, \forall \, k. 
\end{eqnarray}

Similarly, $\bm V^t$ at the $t^{th}$ iteration for problem $\eqref{liftvDC}$ can be obtained by solving the following convex  programming
\begin{eqnarray}\label{subliftvDC}
\mathop{\text{minimize}}_{\bm V}
&&\text{Tr}(\bm V)-\langle \bm{V},\partial_{\bm{V}^{t-1}}\|\bm{V}\|_2\rangle\nonumber \\
\text{subject to}&& \gamma_k^{\text{min}} \left(\sum_{j=k+1}^{K}\text{Tr}(\bm R_{l,j}\bm V) + b_{l,j}^2+ \sigma^2\right) \nonumber\leq \nonumber \\&& \ \ \ \ \ \ \ \ \ \text{Tr}(\bm R_{l,k}\bm V)+ b_{l,k}^2, \forall \, k, l = k,\ldots, K,\nonumber\\
&&\bm V_{n,n} =1 , \forall \, n = 1,\ldots,N+1,\nonumber\\
&&\bm V \succcurlyeq 0.
\end{eqnarray}

Problems $\eqref{subliftomegaDC}$ and $\eqref{subliftvDC}$ are convex and can be efficiently solved  by using CVX.
It is worth noting that the sub-gradient of  $\|\bm X\|_2$ at $ \bm{X}^{t}$ (i.e., $\bm\partial_{\bm{X}^{t}}\|\bm{X}\|_2$) can be efficiently computed as in Proposition 2. 

%for PSD matrix $\bm X\in \mathbb{C}^{N\times N}$
 
\begin{proposition} For PSD matrix $\bm X\in \mathbb{C}^{N\times N}$, the sub-gradient of $\|\bm X\|_2$ can be efficiently computed as 
                \begin{equation*}
      \bm u_1 \bm u_1^{\sf H} \in \bm\partial_{\bm{X}^{t}}\|\bm{X}\|_2,
        \end{equation*}
        where $\bm u_1\in\mathbb{ C}^{N}$ is the eigenvector corresponding to the largest eigenvalue $\sigma_1( \bm{X})$. 
\end{proposition}

The proposed alternating DC algorithm to solve problem $\mathscr{P}_2$ is summarized in Algorithm $\ref{algo2}$, where problems $\eqref{liftomegaDC}$ and $\eqref{liftvDC}$ are solved in an iterative manner until convergence. 
Both DC programming problems $\eqref{liftomegaDC}$ and $\eqref{liftvDC}$ are solved in each iteration by successively solving the convex relaxation of the primal and dual problems of DC programming.
It is worth noting that the proposed alternating DC method can guarantee the feasibility of the rank-one constraint, which yields good network performance.
%In summary, the proposed alternating DC method is to solve problems $\eqref{liftomegaDC}$ and $\eqref{liftvDC}$  in an iterative manner until convergence. The overall algorithm is presented in Algorithm $\ref{algo2}$. 
%Although they are still non-convex, they can be efficiently solved by DC algorithms, which 

\begin{algorithm}[htb]
        \SetKwData{Left}{left}\SetKwData{This}{this}\SetKwData{Up}{up}
        \SetKwInOut{Input}{Input}\SetKwInOut{Output}{output}
        \Input{ Initialize $\bm \Theta^1$ and threshold $\epsilon>0$.}
        \For{$t1=1,2,\ldots$}{
                   Given  $\bm \Theta^{t1}$, solve problem $\eqref{liftomega}$  to obtain the optimal solution $\{\bm W_k^{t1}\}, \forall \, k \in \mathcal{K}$ .\\
                   \For{$t =1,2,\ldots$}{
                                        Select  a subgradient of $\partial \| \bm{W_k}^{t-1}\|_2, \forall \, k \in \mathcal{K}$. \\
                                   Solve convex subproblem $\eqref{subliftomegaDC}$ and obtain the optimal solution $\{\bm W_k^t, \forall \, k \in \mathcal{K}\}$.\\
                          \If{ penalty component of problem $\eqref{liftomegaDC}$ is zero}{\textbf{break}}}
                            Obtain $\{\bm w_k^{t1}\}$ via Cholesky decomposition $\bm W_k^{t} = \bm w_k^{t1}  \bm w_k^{t1\sf H}$. \\
                            
                  Given  $\{\bm W_k^{t1}\}$,  solve problem $\eqref{liftv}$  to obtain  $\bm V^{t1+1}$.\\
                    \For{$t =1,2,\ldots$}{
                        Select  a subgradient of $\partial \|  \bm{V}^{t-1}\|_2$. \\
                             Solve  convex problem $\eqref{subliftvDC}$  and obtain the optimal solution $\bm V^t$.\\
                    \If{ the objective value of  problem $\eqref{liftvDC}$  is  zero}{\textbf{break}}} 
                     Obtain $\tilde{\bm v}^{t1+1}$ via Cholesky decomposition $\bm V^{t} = \tilde{\bm v}^{t1+1}\tilde{  \bm v}^{t1+1\sf H}$. \\  
                     
           \If{ the decrease of the total transmit power is below $\epsilon$ or problem $\eqref {liftv}$ becomes infeasible}{\textbf{break}}}
        \caption{Proposed Alternating DC Algorithm for Solving Problem  $\mathscr{P}_2$.}
        \label{algo2}
\end{algorithm}

\section{Simulation Results}
\begin{figure*}[htb]
        \centering
        \subfigure[Convergence.]{\includegraphics[width=0.5\columnwidth,height=3.5cm]{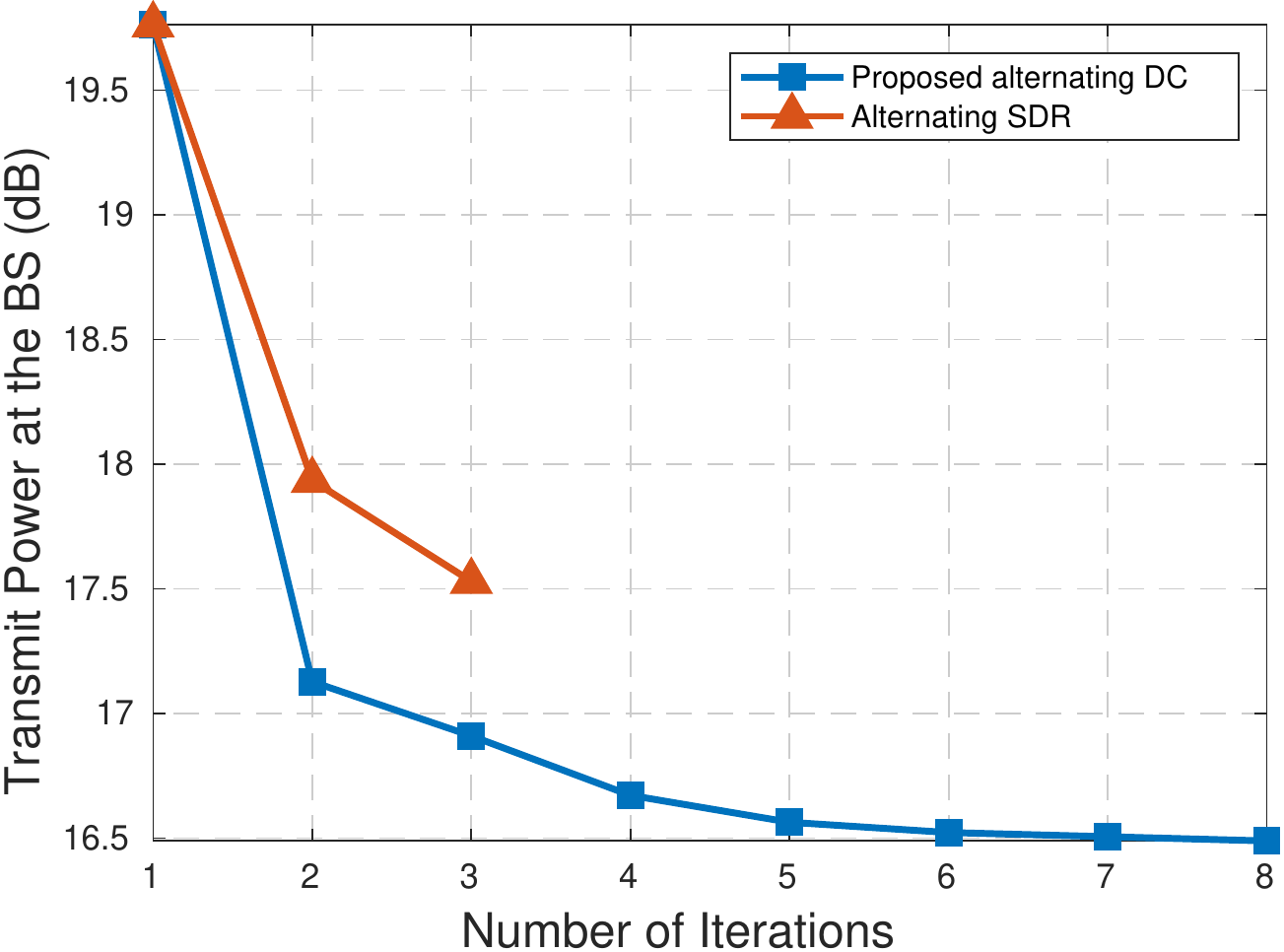}\label{fig:iter}}
        \subfigure[Transmit power vs. $\#$ BS antennas.]{\includegraphics[width=0.5\columnwidth,height=3.5cm]{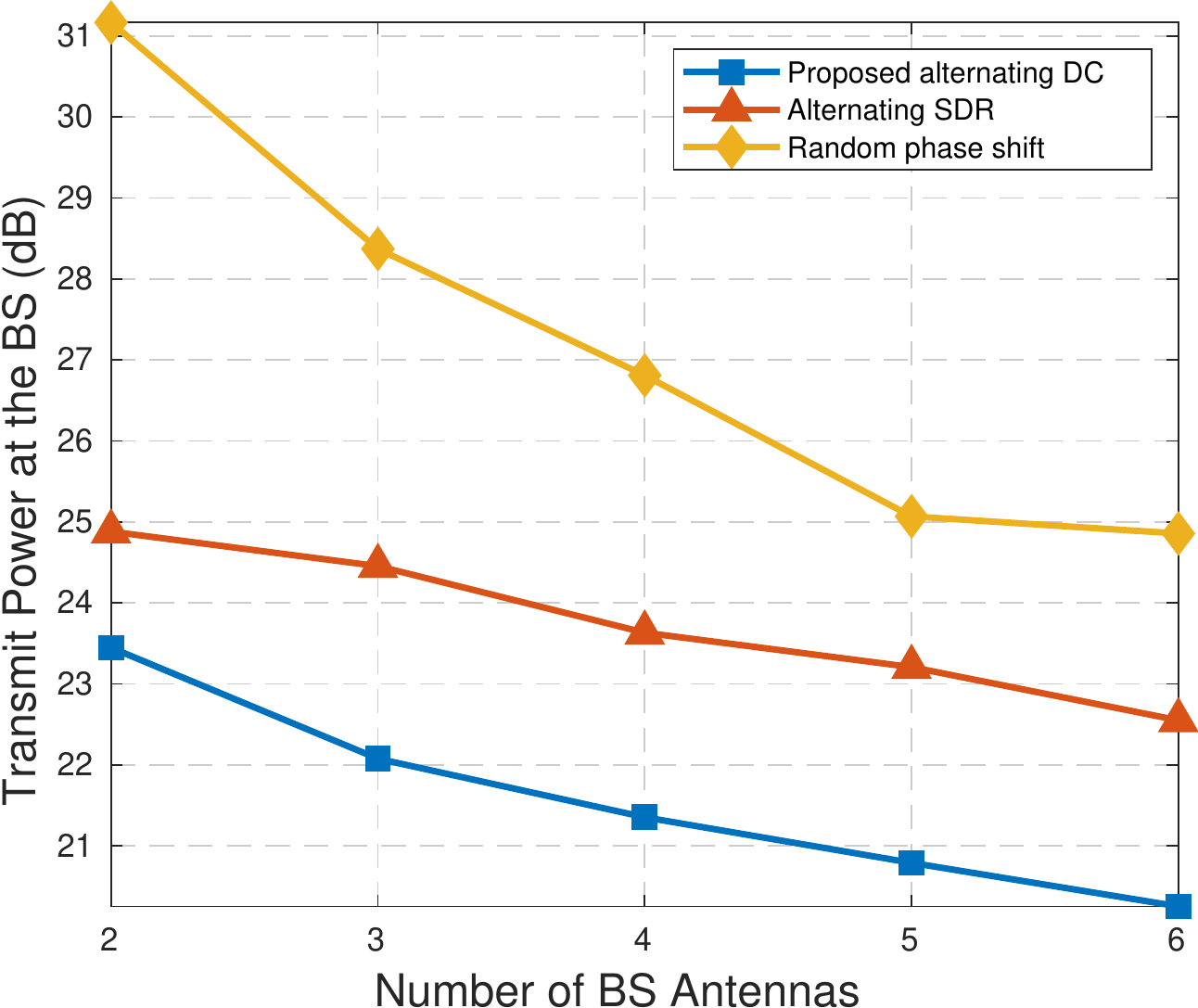}\label{fig:M}}
        \subfigure[Transmit power  vs. \!$\#$ \!IRS elements.]{\includegraphics[width=0.5\columnwidth,height=3.5cm]{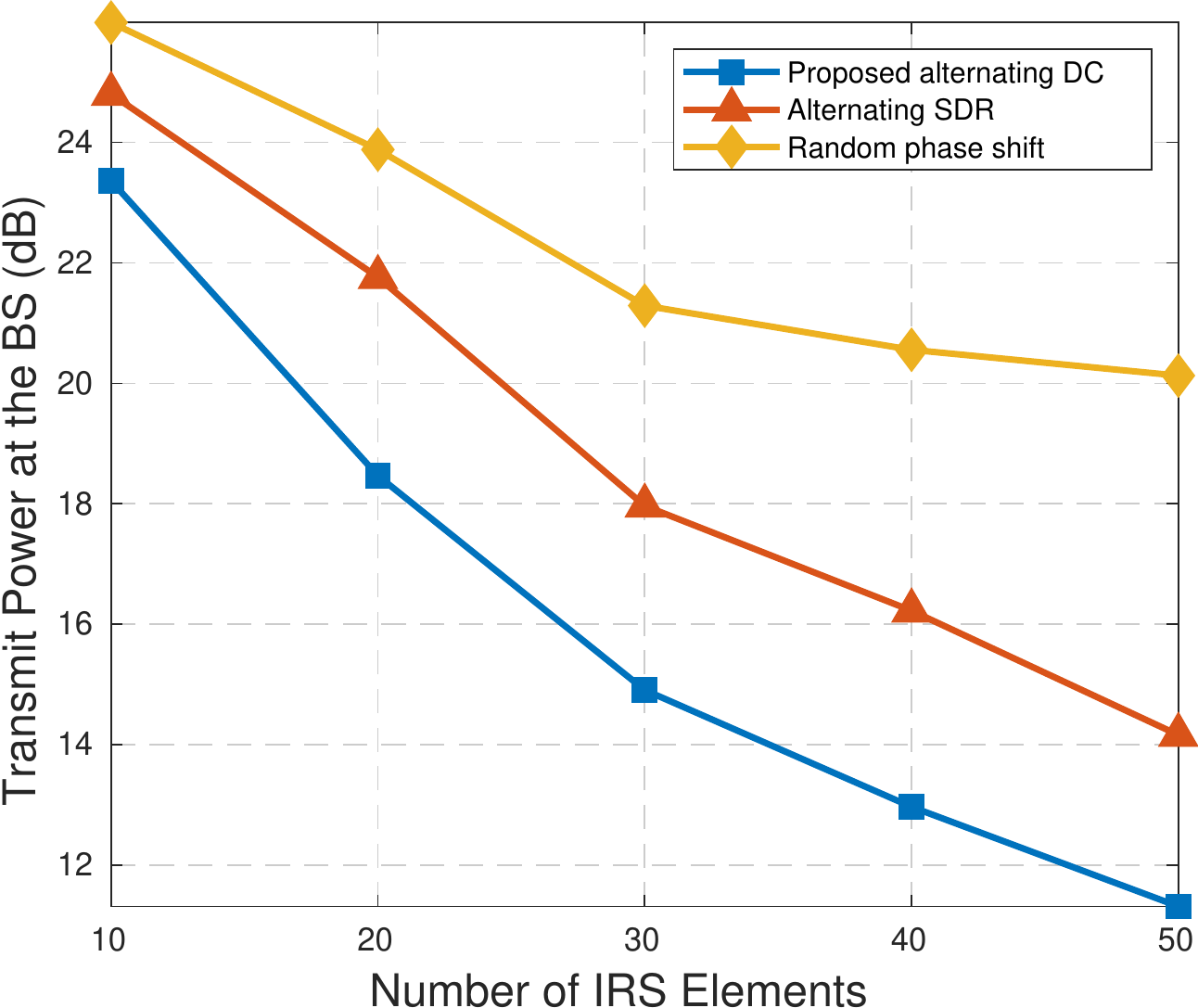}\label{fig:N}}
        \subfigure[Transmit power  vs. $\#$ users.]{\includegraphics[width=0.5\columnwidth,height=3.5cm]{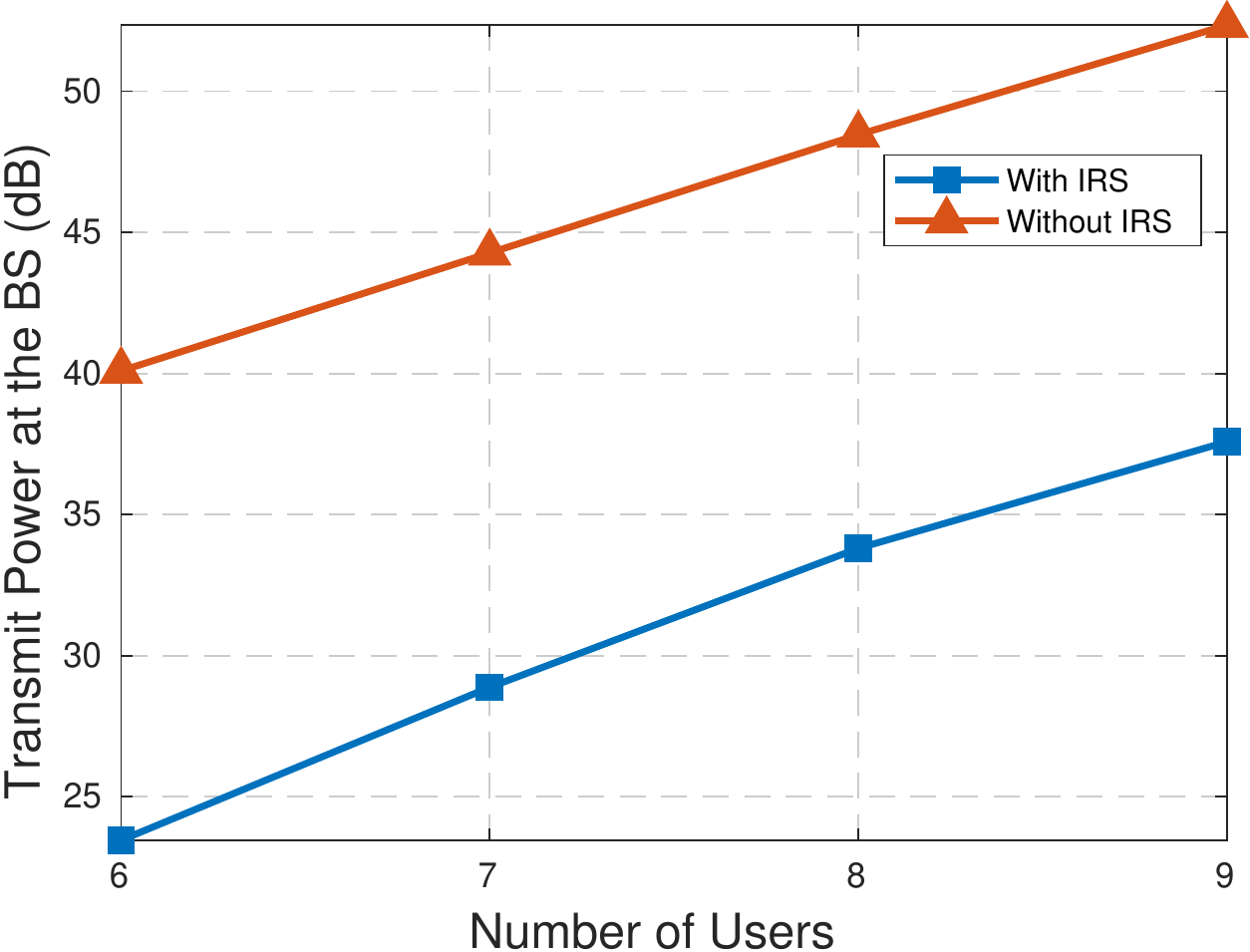}\label{fig:K}}
        \caption{Performance comparisons between the proposed alternating DC method and two baseline methods under different network settings.}\label{fig}
\end{figure*}

In this section, simulation results are presented to demonstrate the effectiveness of the proposed alternating DC method and show the performance of IRS-empowered downlink NOMA networks. 
We consider a three-dimensional (3D) coordinate system, where the BS is located at $(0, 0, 25)$ meters. 
The IRS is placed at $(50, 50, 40)$ meters, where the passive reflecting elements are uniformly distributed on a rectangular surface. 
In addition, the users are uniformly distributed in the region of $(-50,50,0)\times(70,150,0)$ meters. 
The path loss model under consideration is $L(d) = T_0\left({d}/{d_0}\right)^{-\alpha}$,
%\begin{eqnarray}\label{passloss}
%&&L(d) = T_0\left({d}/{d_0}\right)^{-\alpha},
%\end{eqnarray}
where $T_0$ is the path loss at the reference distance $d_0=1$ meter, $d$ is the link distance, and $\alpha$ is the path loss exponent. 
We set $T_0 = 30$ dB, and the path loss exponents for the BS-user link, the BS-IRS link, and the IRS-user link are set to be 3.5, 2.2, and 2.8, respectively. 
All channels are assumed to suffer from Rayleigh fading. 
We denote $d_{BU}^k$, $d_{IU}^k$, and $d_{IB}$ as the distance between user $U_k$ and BS, the distance between user $U_k$ and IRS, and the distance between BS and IRS, respectively. 
Hence, the channel coefficients are given by 
\begin{align}\label{channel}
\bm h_{d,k} = \sqrt{L(d_{BU}^k)}\bm\gamma^d,
\bm h_{r,k} = \sqrt{L(d_{IU}^k)}\bm\gamma^r,
\bm G= \sqrt{L(d_{IB})}\bm\Gamma,\nonumber 
\end{align}
where $\bm \gamma^d\sim \mathcal{CN}(0,\bm I)$, $\bm \gamma^r\sim \mathcal{CN}(0,\bm I)$ and $\bm \Gamma\sim \mathcal{CN}(0,\bm I)$. 
Unless specified otherwise,  we set $R_k^{\text{min}}=1.5, \forall \, k$, $\sigma^2 =0.01$, and $\rho = 20$. 
Each point in Figs.\ref{fig}(b)--(d) is obtained by averaging over 50 channel realizations. 
   
We compare the proposed alternating DC method with the alternating SDR method and the random phase shift method. 
The SDR method solves problems $\eqref{liftomega}$ and $\eqref{liftv}$ alternatively using CVX after removing the rank-one constraints. 
%The algorithm terminates when the difference of the transmit power between two consecutive iterations is below a certain threshold (i.e., $\epsilon$) or the SDR method fails to return a feasible solution to problem $\eqref{liftv}$. 
%%Since the SDR method does not always return a rank-one solution, 
%Gaussian randomization is also applied when the solution obtained by the SDR method does not satisfy the rank-one constraint. 
For the random phase shift method, the phase shift matrix $\bm \Theta$ is randomly chosen and fixed, followed by solving the transmit power minimization problem $\eqref{liftomegaDC}$.

%We also illustrate the results given by random phase shift method as the baseline. That is,  to solve problem $\mathscr{P}_2$,
  % we choose a fixed random phase shift matrix $\Theta$, then minimize the total transmit power by solving problem $\eqref{liftomegaDC}$.

We show the convergence behavior of the proposed alternating DC method and the alternating SDR method in Fig.\ref{fig:iter} when $K = 6$, $M =5$, and $N = 20$. 
It can be observed that the alternating SDR method with Gaussian randomization terminates at the third iteration since it fails to return a feasible solution to problem $\eqref{fixomega}$. In contrast, the proposed alternating DC method is able to induce exact rank-one optimal solutions and hence accurately detect the feasibility of problem $\eqref{fixomega}$. 
  
Fig.\ref{fig:M} shows the impact of the number of BS antennas (i.e., $M$) on the total transmit power when $N=15$ and $K=6$. 
As the value of $M$ increases, the transmit power decreases, which indicates that more antennas at the BS can bring better performance. 
Moreover, both the proposed alternating DC method and the alternating SDR method significantly outperforms the random phase shift method. It demonstrates that jointly optimizing the active beamforming at the BS and the passive phase shifts at the IRS can significantly decrease the transmit power. 
Due to the superiority of the proposed DC representation, the proposed alternating DC method consumes much less transmit power than the alternating SDR method. 

Fig.\ref{fig:N} illustrates the impact of the number of passive reflecting elements at the IRS (i.e., $N$) on the total transmit power when $M=5$, and $K =6$. The total transmit power decreases quickly as the value of $N$ increases, which indicates that a larger number of passive reflecting elements leads to better performance. 
Fig.\ref{fig:K} shows the performance of downlink NOMA networks with and without IRS when $M=5$ and $N =10$. 
The performance of NOMA networks without IRS is obtained by solving problem $\eqref{liftomegaDC}$ with $\bm \Theta = \bm 0$. 
From Fig.\ref{fig:K},  the IRS-empowered networks outperforms  the networks without IRS, which demonstrates the importance of deploying IRS in cellular networks. 

% By increasing the number of users, the BS consumes more transmit power to achieve the target rate. For a specific target data rate,  the two baseline methods consume much more transmit power than the proposed alternating DC method. 
%

\section{Conclusions}
In this paper, we presented an IRS-empowered NOMA method to significantly reduce the transmit power for the emerging 6G networks. To design the beamforming vectors at the BS and the phase shift matrix at the IRS for transmit power minimization, we proposed an alternating DC method. Specifically,
to decouple the beamforming vectors and the phase shift matrix in the formulated
problem, we presented an alternating optimization method by solving two
non-convex QCQP problems alternatively. We then transformed the non-convex QCQP problems into SDP problems via matrix lifting, followed by introducing an exact DC representation for rank-one constraint. Furthermore, we developed an  DC algorithm to solve the resulted DC programming problems. Simulation results demonstrated that the proposed alternating DC method outperforms
 the state-of-the-art methods in terms of the total transmit power in the IRS-empowered NOMA networks.

%\newpage
\bibliography{refs} % BM
\bibliographystyle{IEEEtran}

\end{document}